\documentclass[a4paper]{article}

\usepackage{INTERSPEECH2022}
\usepackage{amsmath,graphicx}
\usepackage{multirow}
\usepackage{amssymb}
\usepackage{booktabs}
\usepackage{cite}

\title{Tandem Multitask Training of Speaker Diarisation and\\ Speech Recognition for Meeting Transcription}
\name{Xianrui Zheng$^*$\thanks{$^*$Supported by an Amazon Studentship.}, Chao Zhang, Philip C. Woodland}
\address{
  Cambridge University Engineering Dept., Trumpington St., Cambridge, CB2 1PZ U.K.
}
\email{\{xz396, cz277, pcw\}@eng.cam.ac.uk}

\begin{document}

\maketitle
\begin{abstract}
Self-supervised-learning-based pre-trained models for speech data, such as Wav2Vec 2.0 (W2V2), have become the backbone of many speech tasks. In this paper, to achieve speaker diarisation and speech recognition using a single model, a tandem multitask training (TMT) method is proposed to fine-tune W2V2. 
For speaker diarisation, the tasks of voice activity detection (VAD) and speaker classification (SC) are required, and connectionist temporal classification (CTC) is used for ASR. The multitask framework implements VAD, SC, and ASR using an early layer, middle layer, and late layer of W2V2, which coincides with the order of segmenting the audio with VAD, clustering the segments based on speaker embeddings, and transcribing each segment with ASR. 
Experimental results on the augmented multi-party (AMI) dataset showed that using different W2V2 layers for VAD, SC, and ASR from the earlier to later layers for TMT not only saves computational cost, but also reduces diarisation error rates (DERs). 
Joint fine-tuning of VAD, SC, and ASR yielded 16\%/17\% relative reductions of DER with manual/automatic segmentation respectively, and consistent reductions in speaker attributed word error rate, compared to the baseline with separately fine-tuned models. 

\end{abstract}

\section{Introduction}
\label{sec:intro}
Multi-party interactions, such as meetings, are natural scenarios for automatic speech recognition (ASR) applications. Since these scenarios often result in long audio streams, prior to applying ASR, it is necessary to obtain individual speech segments by finding ``who spoke when'' using speaker diarisation. 
A speaker diarisation pipeline often consists of at least three modules, namely voice activity detection (VAD), speaker embedding extraction, and clustering, which are normally implemented with separate models \cite{sunCombination2021}. 
A separately trained ASR system can then be used to transcribe each segment found by speaker diarisation, and obtain speaker-attributed ASR output over long audio streams \cite{watanabeCHiME62020,rajIntegration2021}. 
Recently, end-to-end methods have been proposed for jointly modelling some modules in a speaker diarisation pipeline with an ASR system \cite{Kanda2020,horiguchiEndtoend2020,kandaEndtoend2021,Chang2021,LuLiang2021,sklyar2021streaming,sklyar2022multi,kanda2022streaming,khare2022asr}.

Recently, with the progress in self-supervised learning (SSL), speech-based Transformer \cite{vaswaniAttention2017} encoder models pre-trained with SSL \cite{schneiderWav2vec2019,baevskiWav2vec2020,hsuHuBERT2021,chenWavLM2022}, such as Wav2Vec 2.0 (W2V2), have formed the backbone of systems for many speech tasks, including ASR 
and speaker classification (SC) for speaker diarisation \cite{chenWavLM2022}. 
With SSL, a large amount of data without labels is used in pre-training, which can result in general speech representations that are useful for many downstream tasks \cite{baevskiWav2vec2020}. 
It has been found that a reduced amount of labelled data is required when fine-tuning the pre-trained models for a specific downstream task than training from scratch \cite{schneiderWav2vec2019,baevskiWav2vec2020,hsuHuBERT2021,chenWavLM2022}. 
As a result, by sharing the encoder structure obtained by fine-tuning the same pre-trained model for ASR, SC, and other tasks can lead to a simplification of the pipeline for meeting and party transcription tasks. 

In this paper,
a tandem multitask training (TMT)  method is proposed. 
To combine ASR with speaker diarisation, the tasks of VAD, SC, and ASR with connectionist temporal classification (CTC) \cite{gravesConnectionist2006} are achieved by fine-tuning the same W2V2 encoder via multitask training. Only a few fully connected (FC) layers are added to the encoder separately for each task. The proposed multitask training approach for each of the three tasks has FC layers connected to different layers of the encoder. Experimental results on the augmented multi-party (AMI) data for meeting transcription show that, by arranging the three tasks from earlier shared encoder layers to later encoder layers following their order as in the overall processing pipeline that cascades speaker diarisation and ASR, not only reduces diarisation error rates (DERs) but also results in a reduced computational cost. 
Results also demonstrate that using a single shared encoder can provide a better speaker attributed WER on the test sets than fine-tuning a separate encoder for each task.

The paper is organised as follows. Section~\ref{sec:strain} describes separate pipelines for VAD, SC for speaker embedding extraction, and ASR. Section~\ref{sec:mtrain} presents the proposed method that first jointly trains SC and ASR, and later also includes VAD. The experimental setups are given in Sec.~\ref{sec:expsetup}, followed by the results and conclusions in Secs.~\ref{sec:results} and \ref{sec:conclusion}.

\section{Separate training pipelines}
\label{sec:strain}

\subsection{Cascaded speaker diarisation pipelines}
A cascaded pipeline is often used for speaker diarisation, which first finds all active speech regions in the audio stream using VAD, then divides each continuous speech region into multiple fixed-length windows and extracts a speaker embedding for each window, and finally clusters the speaker embeddings and forms the speech segments by joining neighbouring windows from the same cluster. 
The following sub-sections introduce VAD, speaker embedding extraction and ASR, together with the application of the W2V2 encoder for these individual tasks. 

The W2V2 encoder contains three jointly trained components. 
The first component is a feature extractor consisting of a stack of convolutional layers to transform raw waveforms to latent speech representations. The second component is a Transformer \cite{vaswaniAttention2017} based context network, which converts the masked latent speech representations into contextualised representations.
The last component is a quantisation module for discretising the latent speech representations (no mask). 

\subsubsection{VAD for automatic segmentation}

VAD is often employed as the first step for speaker diarisation to divide a long audio stream into speech and non-speech regions. 
Early studies such as \cite{gervenComparative1997} used spectro-temporal properties of speech versus noise. 
Since then, neural network (NN) based VADs have become increasingly popular \cite{woodlandCambridge2015}. 
Multi-layer perceptrons (MLP), long short-term memory (LSTM) and convolutional neural networks (CNN) are common architectures used for VAD. 
The input to the VAD can be a fixed duration of audio \cite{woodlandCambridge2015} and the output is a two dimensional vector indicating the probabilities of speech and non-speech for the current frame at the centre of the input audio. 

In this paper, for VAD a single feed-forward layer is added on top of the W2V2 encoder to project the output from W2V2 to two dimensions and use the softmax activation function to find the probabilities. 
The training objective for VAD is a normal cross entropy loss. 

\subsubsection{SC for speaker embedding extraction and clustering}

After finding the speech regions, normally fixed dimension feature vectors are extracted from each region to represent speakers. 
Examples of the fixed dimensional speaker representations include i-vectors \cite{shumExploiting2011} and more recently d-vectors \cite{varianiDeep2014} and x-vectors \cite{baevskiWav2vec2020,snyderXvectors2018}. 
The d-vectors and x-vectors are obtained from NNs, which are trained on a SC task.
The output from a hidden layer of a SC NN is normally taken as the d-vector/x-vector to represent the speaker information in the diarisation task. 
A pooling layer is often used to convert the frame-level speaker representations to a fixed-length vector for several seconds of audio. Commonly used pooling layers include the means and standard deviations \cite{diezBayesian2019} and those with attention mechanisms \cite{sunCombination2021,shiHvectors2020}

Rather than using the normal cross entropy loss with a softmax activation, some additional loss functions have been proposed to better discriminate between different speaker classes, such as the angular softmax loss \cite{liuLargemargin2017}. 
The additive angular margin loss \cite{dengArcFace2019} is used for all SC training in this paper. 

When fine-tuning the W2V2 encoder for only the SC task, the inputs are fixed length speech segments extracted from utterances. 
The outputs from the encoder are first averaged across time and projected to a lower dimensional space using a FC layer. The averaged features are then mapped into an output space to provide a distribution over the speakers in the training set. 
During evaluation, the output FC layer is discarded, and the encoder with the remaining FC layer is used to provide a speaker embedding for each fixed length speech segment. 
These segments are fed into a clustering algorithm to assign a speaker label to each window. 

\subsection{ASR}

ASR is a popular downstream task for almost all pre-trained speech models. 
Similar to \cite{baevskiWav2vec2020}, only a single FC layer is placed on top of the W2V2 encoder to project the speech representations onto a distribution over the subword units and the CTC \cite{gravesConnectionist2006} loss is used during training.

\subsection{Fine-tuning pre-trained models}
There are two common methods to apply pre-trained models to downstream tasks. 
The first is to update both the pre-trained parameters and the randomly initialised parameters in the newly added layers for the particular task \cite{baevskiWav2vec2020,fanExploring2021}.
The second approach is to freeze all parameters in the pre-trained model and update only the parameters in the newly added layers \cite{yangSUPERB2021}. 
This second method can be more straight forward if the aim is to evaluate the pre-trained model on a large number of tasks at the same time.
With this approach, the learnt representations of the pre-trained models are only used to replace other widely used features such as log mel filterbank. 
However, for some challenging downstream tasks, while the second method often requires more layers to be added after the pre-trained models in order to yield good performance \cite{yangSUPERB2021}, it is often sufficient to add only one FC layer if the first method is used.

\section{Multitask training pipelines}
\label{sec:mtrain}
This section introduces the proposed method of multitask fine-tuning for VAD, SC and ASR with a shared W2V2 encoder. 

\subsection{Jointly modelling SC and ASR}
\label{sec:mtrain_asr_spk}

\begin{figure}[ht]
\centering
\includegraphics{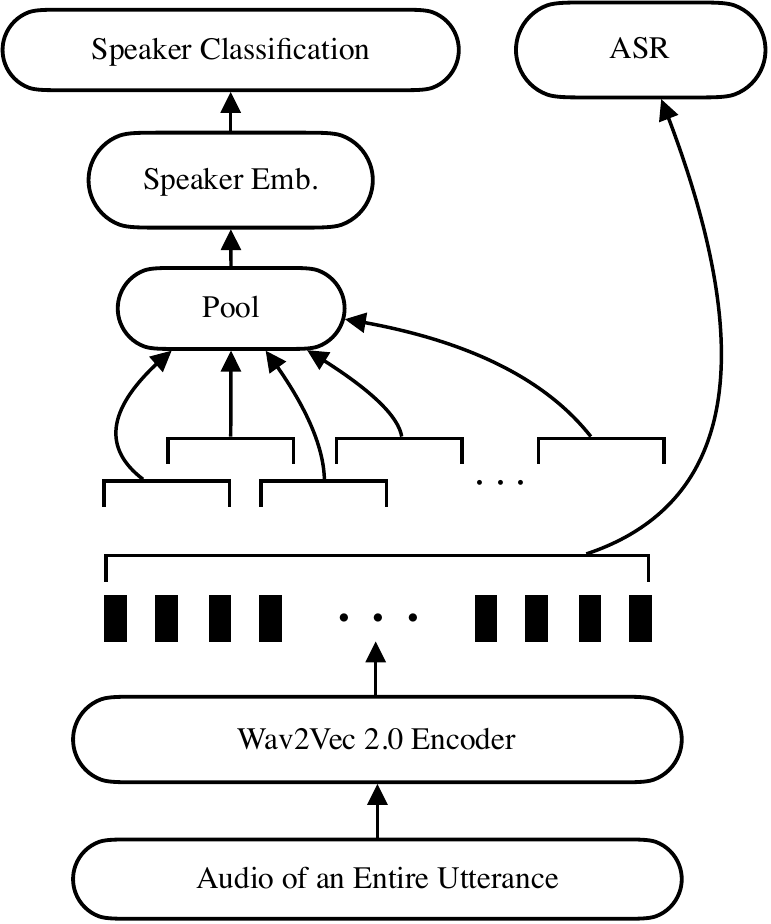}
\caption{Illustration of the proposed multitask fine-tuning for ASR and SC tasks.}
\label{fig:asr_spk}
\end{figure}

When fine-tuning a W2V2 encoder for only the SC task, the input is a segment of a fixed duration audio. 
This input is different from the input required for training an ASR system, which is the audio of an entire utterance. 
To allow these two tasks to be trained together, we modify the pipeline to fine-tune the encoder for the SC task. 
As shown in Fig.~\ref{fig:asr_spk}, the input to the W2V2 encoder is an entire utterance. 
While the outputs of the encoder is mapped to the distribution over the subword recognition vocabulary for the ASR task, these outputs can be sampled using a fixed size window
and then fed into the average pooling layer in parallel before projecting to the speaker embedding dimension and the final distribution over speakers. 
The fixed size window does not need to be applied on the outputs from the same W2V2 encoder layer as the one used for the ASR task. 
Experiments show that using the output from a lower W2V2 layer for the SC task during multitask training yields better diarisation results. 

\subsection{Jointly modelling VAD, SC and ASR}

The training data for VAD is obtained by using a fixed window that steps through the entire meeting. 
These windows can be within an utterance, covering multiple utterances or not covering speech. 
Therefore, it is difficult to use the same input for ASR and VAD. 
The work in \cite{fanExploring2021} chose to sample evenly from two datasets for two different tasks to construct a training batch. 
However, the fixed size window used for VAD is often shorter than the utterances for ASR. 
This can cause redundant memory usage due to the padding required in order for every sample in a batch to have an equal length. 
In order to include VAD in the multitask training framework, we propose training the VAD task at odd training steps, and training ASR and SC tasks at even training steps. 
In this way, all tasks can still share the same W2V2 encoder without sharing the same input waveform. 

\begin{figure}[t]
\centering
\includegraphics{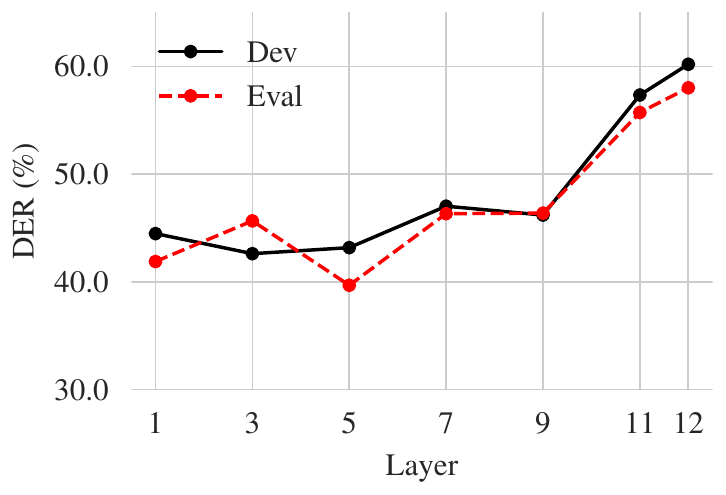}
\vspace{-3mm}
\caption{\%DER on manual segmentation using the pre-trained W2V2 features from different transformer layers without fine-tuning. Overlapped regions are not scored.}
\label{fig:noft}
\end{figure}

\section{Experimental setup}
\label{sec:expsetup}
\subsection{Data}

All models were trained and evaluated on the augmented multi-party interaction (AMI) meeting corpus. 
This corpus contains 100 hours of meeting recordings. 
The training set has 135 meetings with 155 speakers, from which 5\% of the data was used for the validation set. 
The development set (Dev) has 18 meetings with 21 speakers and the evaluation set (Eval) has 16 meetings with 16 speakers. 
Dev has two speakers that are also seen in the training set while all speakers in Eval are unseen. 
The audio data comes from the multiple distance microphone (MDM) data beam-formed by the BeamformIt toolkit \cite{angueraAcoustic2007}. 
The data preparation of the AMI corpus for ASR follows the pipeline provided in ESPnet \cite{watanabeESPnet2018}. 
The same references for speaker diarisation from \cite{sunCombination2021} are used, where the silences in the manual segmentation from the official AMI release were stripped by using a forced alignment with the reference transcriptions and a pre-existing speech recognition system \cite{youngHTK2015}. 
These alignments were also used to generate the speech/nonspeech labels for training the VAD. 

\subsection{Evaluation metrics}

The diarisation error rate (DER) was used as the evaluation metric for measuring the speaker diarisation task, which is the sum of the speaker (clustering) error rate (SER), missed speech (MS) and false alarm (FA). 
The performance of the VAD models are evaluated by MS and FA. 
A 0.25 s collar is set as used for the NIST rich transcription evaluations. 
ASR models were evaluated by using the modified concatenated minimum-permutation word error rate (cpWER) for an unknown number of speakers, which is referred to as cpWER-us. 
The original cpWER was proposed in \cite{watanabeCHiME62020}, where the number of speakers in each session was known to be four. 
Since the actual number of speakers is not known by the speaker diarisation system used in this paper, the number of speakers in the hypothesis may not be the same as that in the reference. 
Therefore, cpWER-us is proposed to accommodate this mismatch. 
When the predicted number of speakers is not the same as that in the reference, the relative speaker IDs in the hypothesis that are not mapped to an absolute speaker ID are removed, and the transcription of a speaker in the reference that are not mapped to a relative speaker is considered to be deletion errors. 

\subsection{Implementation details}

The W2V2 encoder used in this paper is the base version provided by the package `Transformers' \cite{wolfTransformers2020}. 
This encoder is pre-trained on 960 hours of audio from LibriSpeech \cite{panayotovLibrispeech2015} without using the transcriptions and the stack of CNN layers are frozen during fine-tuning. 
The spectral clustering algorithm uses the implementation in SpeechBrain \cite{ravanelliSpeechBrain2021} where the p\_percentile is tuned based on the Dev DER. 
Each input to the VAD contains 3 s audio. 
Since the number of output from the W2V2 encoder is 149 with a 3 s input, each meeting is divided into 3 s segments with 2.98 s stride to  ensure a speech/nonspeech label is given every 20 ms.
The VAD output constrains the time interval between speech regions to be more than 0.04 s, and the post processing step converts any non-speech segment less than 0.4 s to speech. 
For the SC task, the W2V2 features were projected to 128 dimensions to form the speaker embeddings, and then mapped to 155 dimensions corresponding to the number of speakers. 
At the speaker diarisation stage, the fine-tuned model uses the 128-dimensional speaker embeddings given a window of audio to perform spectral clustering.
The maximum and minimum number of speakers for each meeting is set to ten and two respectively.  
The ASR task uses 256 unigram subword units \cite{kudoSubword2018}. 

\section{Results}
\label{sec:results}

\subsection{Without tandem multitask training}

Table~\ref{tab:spk_only} gives the DER results after fine-tuning for the SC task. 
\begin{table}[t]
\begin{center}
\caption{\%DER with manual segmentation using a fine-tuned W2V2 encoder. `L$_n$' means to use the first to the $n$-th W2V2 transformer layers for the task. Overlapped regions are not scored. `Input' refers to the length of the input audio. `Utterance' means the input is the complete utterance.}
\vspace{-3mm}
\begin{tabular}{cc|c|cc}
\toprule[1.5pt]
\multicolumn{2}{c|}{Input} & \multirow{2}{*}{Layers} & \multirow{2}{*}{Dev} & \multirow{2}{*}{Eval} \\ 
Train & Test & & \\
\midrule
\multirow{3}{*}{2 s} & 2 s & L$_{12}$ &
10.7 & 11.3 \\ 
& 2 s & L$_5$ & 
13.5 & 14.2 \\ 
& 3 s & L$_{12}$ & 
10.5 & 10.0 \\ 
\midrule
\multirow{2}{*}{Utterance} & 2 s & L$_{12}$ &
13.8 & 13.7 \\ 
& 3 s & L$_{12}$ &
11.1 & 10.5 \\
\bottomrule
\end{tabular}
\label{tab:spk_only}
\end{center}
\vspace{-2mm}
\end{table}
Without fine-tuning, the lowest DER was achieved using the embeddings from the fifth transformer layer (Fig.~\ref{fig:noft}). 
However, discarding the layers above this layer gave poorer DERs than using all layers after fine-tuning. 
The last two rows in Table~\ref{tab:spk_only} show that the inputs can be the complete utterances for training the SC stage and encoder features can be sampled following Fig.~\ref{fig:asr_spk}. 
The dimension of the window used to sample the output from the W2V2 encoder corresponds to the length of the W2V2 output given a 2 s input audio. 
Although the windows are 2 s during SC training, it is discovered that using a longer window length at diarisation stage yields lower DERs.

The MS and FA of a separately trained VAD is shown in the first line of Table~\ref{tab:allm_vad}. 
The DER results using VAD outputs are presented in Table~\ref{tab:spk_asr_sep_auto}. 
The cpWER-us on manual and VAD segmentations using a separately trained ASR is shown in the last row of Table~\ref{tab:spk_asr_sep_auto}. 
The significant increase in cpWER-us when using automatic segmentation are partly caused by speaker diarisation errors: segments with a incorrect speaker ID have 100\% cpWER-us, and DER is considerably higher when overlapped regions are included during scoring. 
Another difference between using manual and automatic segmentations for cpWER-us is that the former generates output for overlapped regions multiple times while the latter does so only once.

\begin{table}[t]
\begin{center}
\caption{\%DER and \%cpWER-us results on manual and automatic segmentations with separately fine-tuned W2V2 encoders. DERs do not score overlapped regions.}
\vspace{-3mm}
\begin{tabular}{cc|cc|cc}
\toprule[1.5pt]
\multirow{2}{*}{Input (Train)} & \multirow{2}{*}{Metric} & \multicolumn{2}{c|}{Manual} & \multicolumn{2}{c}{Automatic} \\
& & Dev & Eval & Dev & Eval \\ 
\midrule
2 s & \%DER & 10.5 & 10.0 & 14.2 & 12.5 \\ 
Utterance & \%DER & 11.1 & 10.5 & 14.5 & 12.3 \\
\midrule
-- & \%cpWER-us & 31.1 & 32.6 & 45.3 & 43.8 \\
\bottomrule
\end{tabular}
\label{tab:spk_asr_sep_auto}
\end{center}
\end{table}

\subsection{With tandem multitask training}

\begin{table}[t]
\begin{center}
\caption{\%DER scored against the manual segmentation. Numbers inside parentheses scored including overlapped regions. `V', `S' and `A' represent VAD, SC and ASR tasks respectively. }
\vspace{-3mm}
\begin{tabular}{ccc|cc}
\toprule[1.5pt]
\multicolumn{3}{c|}{TMT Tasks} &
\multicolumn{2}{c}{Manual} \\ 
V & S & A & 
Dev & Eval \\
\midrule
-- & -- & -- & 
11.1 (15.2) & 10.5 (13.3) \\
\midrule
-- & L$_{12}$ & L$_{12}$ & 
19.2 (22.5) & 26.1 (28.0) \\
-- & L$_{5}$ & L$_{12}$ & 
8.8 (12.4) & \textbf{7.0} (\textbf{9.7}) \\
\midrule
L$_{12}$ & L$_{12}$ & L$_{12}$ & 
18.1 (21.3) & 22.4 (24.5) \\
L$_{5}$ & L$_{9}$ & L$_{12}$ & 
18.4 (21.5) & 25.9 (27.7) \\
L$_{3}$ & L$_{5}$ & L$_{12}$ & 
12.9 (16.1) & 11.1 (13.6) \\
L$_{1}$ & L$_{3}$ & L$_{12}$ & 
\textbf{8.4} (\textbf{12.0}) & 8.8 (11.4) \\
\bottomrule
\end{tabular}
\label{tab:allm_der}
\end{center}
\end{table}
TMT first integrates the ASR and SC tasks using the proposed method described in Sec.~\ref{sec:mtrain_asr_spk}. 
The window for generating speaker information is 3 s long with a 1 s stride for all experiments shown in Table~\ref{tab:allm_der}. 
The SC+ASR shows that when fine-tuning using L$_{12}$, the last W2V2 encoder layer, for both tasks, the DERs are much higher for the model trained only on the SC task using the last encoder layer. 
This degradation on DER can be resolved by using a lower layer for the SC task during fine-tuning. 
In particular, when the fifth encoder layer is connected to the pooling layer for the SC task while the ASR task still uses the last layer, the DERs on the manual segmentation are significantly better than the model trained only for the SC task.
The relative DER reduction is 21\% and 33\% on Dev and Eval respectively with the manual segmentation. 

The last four rows in Table~\ref{tab:allm_der} includes VAD in the multitask framework. 
Similar to SC+ASR, VAD+SC+ASR also gives very high DERs when all three tasks use the last W2V2 encoder layer. 
Although the DERs of VAD+SC+ASR are lower when the SC task uses the fifth encoder layer, they are still considerably higher than that of the SC+ASR with layers L$_5$ and L$_{12}$. 
This gap can be mitigated by using layers L$_{1}$, L$_{3}$, and L$_{12}$ for VAD+SC+ASR, yielding a DER reduction of 24\%, 16\% on Dev and Eval respectively with manual segmentation over the separately trained model.

\begin{table}[t]
\begin{center}
\caption{VAD results with or without TMT.}
\vspace{-3mm}
\begin{tabular}{ccc|cc|cc}
\toprule[1.5pt]
\multicolumn{3}{c|}{TMT Tasks} 
& \multicolumn{2}{c|}{\%MS} 
& \multicolumn{2}{c}{\%FA}  \\ 
V & S & A & Dev & Eval & Dev & Eval \\
\midrule
-- & -- & -- & 0.4 & 0.7 & 3.7 & 3.1 \\
L$_1$ & L$_3$ & L$_{12}$ & 0.6 & 1.0 & 3.8 & 3.0 \\
\bottomrule
\end{tabular}
\label{tab:allm_vad}
\end{center}
\end{table}

\begin{table}[t]
\begin{center}
\caption{\%DER and \%cpWER-us with automatic segmentation. DERs inside the parentheses scored including overlapped regions.}
\vspace{-3mm}
\begin{tabular}{ccc|cc|cc}
\toprule[1.5pt]
\multicolumn{3}{c|}{TMT Tasks} & \multicolumn{2}{c|}{\%DER} & \multicolumn{2}{c}{\%cpWER-us} \\ 
V & S & A & Dev & Eval & Dev & Eval \\
\midrule
-- & -- & -- & 14.5 (17.9) & 12.3 (14.8) & 45.3 & 43.8 \\
-- & L$_5$ & L$_{12}$ & \textbf{11.5} (\textbf{14.8}) & 11.0 (13.4) & \textbf{42.5} & 43.3 \\
L$_1$ & L$_3$ & L$_{12}$ & 11.7 (15.1) & \textbf{10.2} (\textbf{12.6}) & 42.9 & \textbf{42.3} \\
\bottomrule
\end{tabular}
\label{tab:allm_der_asr}
\end{center}
\end{table}

Table~\ref{tab:allm_vad} compares the model trained for VAD only and VAD+SC+ASR. 
Even though VAD+SC+ASR only uses the first layer of the W2V2 encoder for the VAD task, the performance is similar to the model without multitask training using the last layer of the W2V2 encoder. 

Automatic segmentation was used to find the DER and cpWER-us in Table~\ref{tab:allm_der_asr}. 
For the first two lines, a separately trained VAD was used to provide segmentations. 
The last line uses the VAD jointly trained with the other two tasks. 
Both SC+ASR and VAD+SC+ASR performed better than models trained from scratch on Eval. 
SC+ASR gives relative reductions of 11\% and 1\% for Eval DER and cpWER-us respectively, and
VAD+SC+ASR gives relative reductions of 17\% and 3\% for Eval DER and cpWER-us respectively.

\section{Conclusion}
\label{sec:conclusion}
This paper has investigated the use of a W2V2 encoder for speaker diarisation and ASR. A tandem multitask training (TMT) method has been proposed that jointly trains VAD, SC, and CTC-based ASR based on different W2V2 layers. The proposed method was evaluated for meeting transcription on the AMI dataset. When combining SC and ASR based with manual segmentation, TMT resulted in a 33\% reduction in speaker diarisation error rate (DER) on the AMI evaluation set. 
The goal of combining speaker diarisation with ASR was achieved by also incorporating VAD into the TMT approach. 
Results showed the jointly trained model with VAD, SC, and ASR yielded a 17\% relative reduction in DER over the separately trained models and can save computation and storage costs due to the use of a shared encoder.

\bibliographystyle{IEEEtran}
\bibliography{mybib}

\end{document}